\documentstyle{fbssuppl}
\title{Reworking the Tucson-Melbourne Three-Nucleon Potential
\footnote{Dedicated to the 60th birthday of Walter Gl\"ockle.}}
\author{S. A. Coon and H. K. Han  }
\instlist{
Physics Department, New Mexico State University, Las Cruces, NM 88003,
U.S.A.}

\begin{document}
\maketitle

\begin{abstract}
We introduce new values of the strength constants 
(i.e., $a$, $b$, $c$, and $d$ coefficients) of the Tucson-Melbourne (TM)
2$\pi$ exchange three nucleon potential.  The new values come from 
contemporary dispersion relation analyses of meson
factory $\pi$N scattering data.  We make variational Monte Carlo
calculations of the triton with the original and updated three-body
forces to study the effects of this update.  We remove
 a short-range -- $\pi$-range part of the potential due to the $c$
coefficient  and discuss the effect on the triton binding energy.

\end{abstract}

\section{Introduction}

The Tucson-Melbourne (TM) three-nucleon force due to two-pion exchange
has a structure which, after an expansion of the invariant $\pi$N
amplitudes in the inverse nucleon mass,  was determined by the original
implementation of chiral symmetry in the underlying $\pi$N scattering
amplitude.  Given that structure, the  strength
constants (the $a$, $b$, $c$, and $d$ coefficients) are then not free
parameters but depend upon the  $\pi$N scattering data base, which has
improved greatly since the original determination of these
coefficients.  In this note, we review two recent developments in
three-body force studies: i) a critical analysis of the generic
structure of  a 2$\pi$ exchange three-body force (TBF)~\cite{FHvK_99},
and ii) the new TM strength constants derived from  invariant $\pi$N
amplitudes~\cite{KH} corresponding to the contemporary data base which includes
measurements taken at the meson factories since 1980.  We make updated
TBF's of the Tucson-Melbourne type which reflect one or both developments, 
add them to a NN force, and  calculate 
properties of the triton in order to see the effect of these
developments in a simple nuclear system.  

To begin, we display the Tucson-Melbourne force (leaving out an overall
momentum conserving delta function):

\begin{eqnarray}
\lefteqn{
\langle {\vec {p'}_1} {\vec {p'}_2} {\vec {p'}_3} | W_{\pi\pi}(3) |
        {\vec p}_1 {\vec p}_2 {\vec p}_3 \rangle   = } \nonumber \\
& &  (2\pi)^3
\frac{(\vec {\sigma}_1 \cdot {\vec q}) (\vec {\sigma}_2 \cdot {\vec {q'}}) }
     { ({\vec q}^2 + \mu^2) ({\vec {q'}^2} + \mu^2) }
\frac{g^2}{4m^2}
F^2_{\pi NN}({\vec q}\,^2)
F^2_{\pi NN}({\vec {q'}}^2)
\nonumber \\
& & \left\{ ( \vec{\tau}_1 \cdot \vec {\tau}_2 )
     \left[ a + b {\vec q} \cdot {\vec {q'}}
           +c({\vec {q}\,^2} + {\vec {q'}}^2) \right]
+(i \vec {\tau}_3 \cdot \vec {\tau}_1 \times \vec {\tau}_2)
d
(i \vec {\sigma}_3 \cdot {\vec q} \times {\vec {q'}}) \right\}
\, , \label{eq:wpipi} \nonumber\\
\end{eqnarray}
where $\vec {q} = {\vec p}_2 - {\vec {p'}_2}$ and  $\vec {q'} = {\vec
p'}_3 - {\vec {p}_3}$ and the pion rescatters from nucleon 3. (We refer
the reader to Refs.~\cite{CG_81,CP_93} for  diagrams, more extensive
definitions,  explanations of the other two cyclic terms, etc. needed
for calculation but  not directly relevant to the present discussion). 
Now we  review briefly the  origin of this equation.

 The approach used in the Tucson-Melbourne (TM) family of forces is
based upon applying the Ward identities of current algebra to
axial-vector nucleon scattering. The Ward identities are saturated with
nucleon and $\Delta(1230)$ poles. Then employing PCAC (partial
conservation of the axial-vector current), one can derive expressions
for the on-mass-shell pion-nucleon scattering amplitudes~\cite{ST}
which map out satisfactorily the empirical coefficients of the
H\"{o}hler subthreshold crossing symmetric expansion based on
dispersion relations \cite{hohlerbook} and, after projection onto
partial waves, describe the phase shifts reasonably well~\cite{MC_95}.
 The off-mass-shell extrapolation (needed for the exchange of virtual,
spacelike pions in a nuclear force diagram) is trivial for the $d$
coefficient.  It can be taken directly from the  on-mass-shell
theoretical or empirical amplitude $\bar{B}^-$ since they coincide so
closely (see Appendix A of Ref.~\cite{Coon_79}). One can    treat this
coefficient more elaborately~\cite{CG_81,CP_93}, but the result is the
same.  On the other hand, one really needs an off-shell $\pi N$
amplitude for the important    $a$, $b$, $c$ structure of
 Eq. (\ref{eq:wpipi}).    This structure relies on the fact that
the off-pion-mass-shell amplitude $\bar{F}^+$ can be written in a form
which depends on measured on-shell amplitudes only.  This rewriting of
the PCAC/current algebra amplitude exploits a convenient correspondence
between the structure of the terms corresponding to spontaneously
broken chiral symmetry and the structure of the model $\Delta$ term. To
see this, we note that the nonspin flip $t$-channel isospin even
amplitude (covariant nucleon pole term removed) is 
\begin{equation}
  \bar F^{+}(\nu ,t,q^{2},q'^{2}) = f(\nu ,t,q^{2},q'^{2}) \frac{\sigma}
    {{f_{\pi}}^2}
      + C^{+}(\nu ,t,q^{2},q'^{2})
 \label{eq:fampli}
\end{equation}
where $\sigma$ is the pion-nucleon $\sigma$ term,  $f_{\pi} \approx 93$
MeV, and the  invariant amplitude $\bar F^{+}(\nu, t)$ 
is  given in units of the charged pion mass (139.6 MeV).
The double divergence $q'\cdot \bar{M}^+\cdot q/f^2_{\pi}$ of
the background axial vector amplitude denoted by $C^{+}$
contains the higher order $\Delta$ isobar contribution.  
In general, $C^+$ must have the simple form \cite{ST,Coon_79}
\begin{equation}
   C^{+}(\nu ,t,q^{2},q'^{2})=c_{1}{{\nu}^{2}}+c_{2} q \cdot q' +O(q^{4})
\, .
\label{eq:cexpa}
\end{equation}
On the other hand,
the assumed form of the function $f$,
\begin{equation}
     f(\nu ,t,q^{2},q'^{2})=(1-\beta)(\frac{q^{2}+q'^{2}}{{m^{2}_{\pi^+}}}-1)+
\beta (\frac{t}{{m^{2}_{\pi^+}}}-1)
\label{eq:funf}
\end{equation}
(adapted \cite{Coon_79,Praguelec} for $\pi$N scattering from the $SU(3)$
generalization of the Weinberg low energy expansion for $\pi\pi$
scattering )
is such that $\bar F^{+}$  satisfies  soft pion theorems (for a review
see Ref.~\cite{Praguelec}),
and (with the aid of Eq. (\ref{eq:cexpa})) the constraint
at the (on-shell and measurable)  Cheng-Dashen point:
\begin{equation}
\bar F^{+}(0, 2 {m^2_{\pi^+}},{m^2_{\pi^+}},{m^2_{\pi^+}})= 
\frac{\sigma}{{f_{\pi}^2}} + {\cal O}(q^4)\,.
\label{eq:chengd}
\end{equation}
The value of $\beta$ can be determined by taking the amplitude on-shell
and comparing with on-shell data
extrapolated into the subthreshold region~\cite{ST,Praguelec}, but it is
not needed, as we will now demonstrate.

Neglecting the $\nu^2$ and $q_0$ terms in (\ref{eq:fampli}) because they
are of the order of $(m_{\pi^+}/m)^{2}$ or higher,
 the $\bar F^{+}$ amplitude can be expanded in
the three-vector pion momenta $\vec{q}$ and $\vec{q'}$ as
follows:
\begin{equation}
\bar F^{+}(0,t,q^2,q'^2) = -\frac{\sigma}{{f_{\pi}}^2}
+(\frac{\sigma}{{f_{\pi}}^2} \frac{2 \beta}{m_{\pi^+}^2} - c_2) \vec{q} \cdot
\vec{q'}
-\frac{\sigma}{m_{\pi^+}^2 {f_{\pi}}^2} (\vec{q}\,^2 + \vec{q'}^2)
\label{eq:expan}
\end{equation}

The last equation explicitly exhibits the separation between
the (higher order in $\vec{q}\,^2$) $\Delta$ contribution
--- contained in the $c_2$ term alone ---
and the remaining chiral symmetry breaking terms. In ref.
\cite{Coon_79}
and subsequent discussions of the TM $\pi-\pi$ force,
the $c_2$ and $\beta$ constants in
the coefficient of the $\vec q \cdot \vec {q'}$
term were eliminated in favor
of the on-shell (measurable) quantity $\bar F^{+}(0,m_{\pi^+}^2,m_{\pi^+}^2,
m_{\pi^+}^2)$
\begin{equation}
\bar F^{+}(0,m_{\pi^+}^2,m_{\pi^+}^2,m_{\pi^+}^2)
= (1-\beta)\frac{\sigma}{{f_{\pi}}^2} +\frac{c_2 m_{\pi^+}^2}{2}
\label{eq:fonshell}
\end{equation}

From the expanded $\pi$N amplitude $\bar F^{+}$ in conjunction
 with the $\pi$NN vertices $F_{\pi NN}({\vec q}\,^2)$ and
pion propagators, one constructs the three body force of Eq.
(\ref{eq:wpipi}).  Comparing Eqs. (\ref{eq:wpipi}) and (\ref{eq:expan}),
 ($W \sim T$ and $(S-1) =-iT$  so that $T = -F$~\cite{CP_93}) one sees that 
\begin{equation} 
	a = +\frac{\sigma}{{f_{\pi}}^2}\; .  \label{acoef}
\end{equation}	
The $\Delta$ constant $c_2$ contributes then to the overall
coefficient ``b" that has been used in nuclear calculations
($b= b_{\sigma} + b_{\Delta};\,\, b_{\Delta}=c_2$)
\begin{equation}
b = -\frac{\sigma}{{f_{\pi}}^2} \frac{2 \beta}{m_{\pi^+}^2} + c_2 =
-\frac{2}{m^2_{\pi^+}}\left[ \frac{\sigma}{{f_{\pi}}^2}
- \bar F^{+}(0,m_{\pi^+}^2,m_{\pi^+}^2, m_{\pi^+}^2)\right]
\label{bcoef}
\end{equation}
Finally the $c$-term of Eq. (\ref{eq:wpipi}) is given by
\begin{equation}
c = \frac{\sigma}{m_{\pi^+}^2 {f_{\pi}}^2}- \frac{g^2}{4m^3}
 + F'_{\pi NN}(0) \frac{\sigma}{{f_{\pi}}^2}
\label{ccoef}
\end{equation}
The dominant part of $c$ comes from our ansatz Eq.(\ref{eq:funf}) but a
small part is due to the 
 backward-propagating nucleon term $ F^+_Z$ (``Z-graph") 
 $F^+_Z ={\textstyle \frac{g^2}{4m^3}}(\vec{q}\,^2 + \vec{q'}^2)$.  This
 term (which also appears in the $d$ coefficient) is representation
 dependent and is the only local term of a consistent set of 15 terms
 derived some time ago~\cite{CF_86}.
We note that the term proportional to $ F'_{\pi NN}(0)$ did not
appear before in Eq. (\ref{eq:expan}).
This term nevertheless is inserted in $c$ because both the
backward-propagating part
of the nucleon pole $F^{+}_Z$ and the $\Delta$  couple with the
pion with a (assumed the same) form factor $ F_{\pi NN}(q^2)$ which
is defined as $g(q^2) = g  F_{\pi NN}(q^2)$.
The chiral breaking $\sigma$ term 
  has no intrinsic $q^2$ dependence (although
it is multiplied by $f(\nu, t, q^2, q'^2)$). It is convenient, if not
necessary, however, since part of the amplitude is due to $F^{+}_Z$
and $C^{+}$, to multiply the final amplitude by form factors, dependent
upon $q^2$ and $q'^{2}$. Consequently, the constant term
($\sigma/{f_{\pi}}^2$, labeled ``a" in the literature) attains a
spurious momentum dependence
from the form factors. The term proportional to $ F'_{\pi NN}(0)$ in
Eq. (\ref{ccoef}) is inserted to correct for this
spurious momentum dependence to the orders in $q^2$ and $q'^2$
kept in the amplitude.

The new development in the structure of a 2$\pi$ exchange
TBF~\cite{FHvK_99} lies in another look at the decomposition of the
$c$-term
made originally~\cite{Coon_79} to  Fourier
transform  Eq. (\ref{eq:wpipi}), but true in general.
Begin with  the schematic structure 
\begin{equation}
W(3)|_c \propto
{1\over {\vec q}^2+\mu^2}\ {1\over {\vec {q'}}^2+\mu^2}\ 
              ({\vec q}^2 + {\vec {q'}}^2)\ \vec \tau_1 \cdot \vec
	      \tau_2\\    \label{wc}
\end{equation}	      
and rewrite it (neglecting the isospin dependence in Eq.~(\ref{wc})) as
\begin{eqnarray}
\label{decomp}
W(3)|_c&\propto&
{\vec q^2\over \vec q^2+\mu^2}\
{1\over \vec {q'}^2+\mu^2}+(q\leftrightarrow q')\nonumber\\
&=&{\vec q^2+\mu^2-\mu^2\over \vec q^2+\mu^2}\
{1\over \vec {q'}^2+\mu^2}+(q\leftrightarrow q')\nonumber\\
&=&\left( \underbrace{1}_{SR}-
\underbrace{\mu^2\over \vec q^2+\mu^2}_{\pi{\rm -range}}\right) \
{1\over \vec {q'}^2+\mu^2}+(q\leftrightarrow q')
\end{eqnarray}
Thus the $c$-term can be decomposed into a $2\pi$-exchange term with
the same  operator structure as the $a$-term plus a short-range --
$\pi$-range term. Without a form factor $F_{\pi NN}(\vec{q}^2)$ the
short-range part would be a Dirac delta function--a zero-range or
contact term.  This operator structure is reflected in the coordinate
space representations where one always finds the coefficient
$a-2\mu^2c$ multiplying derivatives of two ``coordinate space Yukawas":
see, for example, Eqs. 3.9-3.11 of Ref.~\cite{Coon_79}	or Appendix A
of Ref.~\cite{CPW_83}. Without a form factor $F_{\pi NN}(\vec{q}^2)$
the short-range part would be a Dirac delta function--a zero-range or
contact term. The Tucson-Melbourne force has an (unadorned by $\mu^2$) 
$c$      coefficient multiplying a derivative of a product of a delta
function and a ``coordinate space Yukawa" as is easily seen in the
same equations.

It was the latter, rather singular, aspect of the Tucson-Melbourne
force which made numerical work difficult in both coordinate space and
momentum space (the operator structure is the same). In addition, the
recent trend toward  a low mass cutoff $\Lambda$ in 
\begin{equation}
F_{\pi NN}(\vec{q}^2) = \frac{\Lambda^2 - \mu^2}{\Lambda^2 + \vec{q}^2}
\end{equation}	
for pion exchange highlights the
point already emphasized by the Hokkaido group \cite{Sapporo} and, in
the modern context, by the S\~{a}o Paulo group \cite{contact}.  The
contact terms (those proportional to a coordinate-space
$\delta$-function and its derivatives) are spread out with increasing
importance as $\Lambda$ becomes smaller and the (strong interaction)
size of the nucleon grows.  These groups contended  that these contact
terms, bringing the nucleon structure signature, should not be included
in potential models.

The subject of contact terms has been revived recently with the advent
of effective field theories in which contact terms are used to  emulate
the short distance physics, and the long distance physics, including the
physics of  chiral symmetry, is retained explicitly.  In these
effective field theories (chiral perturbation theory extended to two or
more nucleons~\cite{Bira}) contact terms abound, both in the chosen
chiral Lagrangian and in the nucleon potentials.  Adapting a field
redefinition technique first used in pion condensation~\cite{TW_95}, 
Friar {\em et al.}~\cite{FHvK_99} were able to demonstrate, via a field
theoretic calculation with  an effective chiral Lagrangian,  why the
contact term of Eq. (\ref{decomp}) does not appear in the $2\pi$-three
body force of chiral perturbation theory, even though that field theory
can be transformed to emulate the soft pion theorems.  In sum, although
chiral symmetry in the form of PCAC/current algebra motivated the
ansatz Eq. (\ref{eq:funf}) which led to the operator structure of Eq.
(\ref{decomp}), chiral symmetry in the form of effective field theory
dictates that only the $2\pi$-exchange part ($\propto a - 2\mu^2c$ for
TM) should be retained in a TBF from pion exchange.  One moral which
can be drawn from this new insight is that chiral constraints on the
off-shell scattering amplitude are not enough to determine a
three-nucleon force; one must also satisfy chiral constraints on the
on-shell three-nucleon $S$-matrix elements which are presumed to make
up the force.  This observation applies to other off-shell amplitudes
embedded in nuclear force models~\cite{CMR_97}.

 The removal of the spurious contact term from the Tucson-Melbourne
force leaves a TBF with coefficients $a' = a - 2\mu^2c$, $b$, and $d$
which has been termed TM$'$ in Ref.~\cite{FHvK_99} and subsequent works. 
In the following section we will examine the effects in the triton of
the original TM TBF and the TM$'$ TBF.  We consider TM and TM$'$  with
 the original strength constants and with strength constants from the
current $\pi$N scattering data.

\section{Numerical Results}
	
	We employ  a variational Monte Carlo method developed for
accurate numerical calculations of light nuclei~\cite{CPW_83}.  
 The ``Urbana-type   potentials", suited
to this variational approach, take the form of  a sum of operators
multiplied by functions of the interparticle distance. Following our
previous study of charge symmetry breaking in light
hypernuclei~\cite{Prohunice}, and in order to compare with other TBF
studies~\cite{LANL_88,Han_95}, we use the
  Reid soft core nucleon-nucleon potential in
the form of the Urbana-type Reid $v_8$ potential~\cite{reidv8}. 
The Reid $v_8$ is a simplified (the sum of operators is truncated from
a possible 18~\cite{AV18} to  8 operators) $NN$ force model which is equivalent to
the original Reid soft core nucleon-nucleon potential in the lower
partial waves and can produce the dominant correlations in s-shell
nuclei.  To be specific, the Reid $v_8$ is obtained from the Reid soft
core (RSC) potential in the singlet states $^1S_0$ and $^1P_1$
and the triplet states $^3S_1-^3D_1$ and $^3P_2-^3F_2$.  
   The binding
energy of the triton, calculated with exact Faddeev codes which include
all partial waves $j \leq 4$ (34 channels), is -7.59 MeV for
the Reid $v_8$ (as quoted in Table IV of \cite{APW_95}), to be compared
with -7.35 MeV obtained with the original RSC \cite{LANL_88}. 
This small discrepancy, presumably due to differences in the $P$-waves
of the two  potentials, should not affect our conclusions.

The variational method we use, with Monte Carlo evaluations of the
integrals, is described in Ref. ~\cite{cwbook} (see also,
Ref.~\cite{Wir_91}). Here we
specify only the {\em differences} from the equations in these
references.  In particular,
the trial nuclear wave functions have the following 
structure:
\begin{equation}
\Psi={\bf S}\left[\prod^{A}_{i<j<k}f_{ijk}\right]
{\bf S}\left[\prod^{A}_{i<j}f_{ij}\right]\Phi,
\end{equation}
where $\Phi$ is an antisymmetric spin-isospin state,
 having appropriate values of
total spin and isospin, with no spatial dependence, and $\bf S$ is a
symmetrization operator which makes 3! terms for the two-body
correlation operator  $f_{ij}$ and one term for the 
three-body correlation operator $f_{ijk}$.  
The NN correlation operator is
\begin{equation}
f_{ij}=f_{ij}^c+f_{ij}^\tau \mbox{\boldmath $\tau$}_i
 \cdot \mbox{\boldmath $\tau$}_j
+f_{ij}^\sigma \mbox{\boldmath $\sigma$}_i\cdot \mbox{\boldmath
$\sigma$}_j
+f_{ij}^{\sigma \tau} \mbox{\boldmath $\sigma$}_i\cdot \mbox{\boldmath
$\sigma$}_j \mbox{\boldmath $\tau$}_i
 \cdot \mbox{\boldmath $\tau$}_j
+f_{ij}^t S_{ij} +f_{ij}^{t \tau}S_{ij}\mbox{\boldmath $\tau$}_i
\cdot \mbox{\boldmath $\tau$}_j
\end{equation}
and the triplet correlation induced by the three-body force has the usual 
  linear form  suggested by the first order perturbation 
theory \cite{Carlson III}
\begin{equation}
f_{ijk}=1+\beta V_{ijk} 
\end{equation}
where $\beta$ is a  variational parameter.
These pair correlations 
 do not include the spin-orbit correlations described in
Ref.~\cite{Wir_91}, nor do our triplet correlations include the more
sophisticated three-body correlations introduced by Arriaga {\em et
al.}~\cite{APW_95} which reduce the difference between the  variational
upper bound and the Faddeev binding energy of the triton to less than
2\%.  Both improvements would be clearly desirable, but are beyond the
scope of this preliminary investigation.  We do, however,  include the
  usual central three-body correlation $f_3$ multiplied 
by the correlation functions($f_{ij}^\tau$, 
$f_{ij}^\sigma$, $f_{ij}^{\sigma \tau}$, $f_{ij}^t$ and 
$f_{ij}^{t \tau}$):
\begin{equation}
f_3=\prod_{k\ne i,j}^3\left[ 1-t_1\left(\frac{r_{ij}}{R}
\right)^{t_2} {\rm exp}(-t_3 R) \right]
\end{equation}
with $R=r_{ij}+r_{ik}+r_{jk}$.  With these correlations we  get a
binding energy of -7.28(3) MeV with the Reid $v_8$ alone, a number
which compares favorably with variational results in Table V of
Ref.~\cite{Wir_91},  obtained with a slightly different trial wave
function.  We now demonstrate that our variational calculations track
the Faddeev results of Ref.~\cite{LANL_88} and suggest that the main
outline of our results (to be presented later) will reflect the
properties of the Hamiltonians chosen, provided that the potentials
are not too singular.

The Faddeev calculations we now examine used the RSC potential and the
early parameters (labeled TM(81) here) of the Tucson-Melbourne TBF
($m_{\pi^+}$a = +1.13, $m^3_{\pi^+}$b = -2.58, $m^3_{\pi^+}$c = 1.00, 
and $m^3_{\pi^+}$d =
-0.753  in units of the charged pion mass: 139.6 MeV) obtained from an
interior dispersion relation (IDR) analysis of phase shifts circa
1973~\cite{HJS_76}.  Ten years ago there was little reason to look
suspiciously at the $c$-term, and the goal of the exercise was to test
the perturbative nature of the $\pi N$ amplitude $s$-wave terms.  To
this end, a restricted model  was chosen with $a=c=0$, the Faddeev
eigenvalues calculated  for a 34-channel solution for RSC/TM, and the
solution tested by employing the resulting wave functions in a
Raleigh-Ritz variational calculation.  The variational result for this
restricted Hamiltonian coincided with the Faddeev eigenvalue,
indicating the high quality of the Faddeev wave function.  Then the $a$
and $c$ terms were selectively set to their assigned values and the
variational calculation was repeated.  Comparison of the results shows
the non-perturbative role of the $a$ and the $c$ term on the triton
wave function.  To test our codes and to suggest that our methods can
give insight into triton binding energy effects from the proposed
redefinitions of the TM force, we made a parallel set of calculations
with the Reid $v_8$ and the old TM force, TM(81), with the parameters
given above.  The results are shown in table 1.

\begin{table}[hbt]
\begin{center}
\begin{tabular}{lcc}
        &RSC/TM(81)&Reid $v_8$/TM(81)\\  \hline

$-\langle H_{a=0,c=0}\rangle$&   9.18 & 9.12\\
$-\langle H_{c=0}\rangle$& 9.07& 8.99 \\
$-\langle H_{a=0}\rangle$& 8.46 & 8.17 \\
$-\langle H\rangle$& 8.35 & 8.04 \\

\end{tabular}  \\
\caption{Hamiltonian expectation values for RSC plus
TM variations indicated and Reid $v_8$ plus TM variations indicated.}
\end{center}
\end{table}

From Table 1, we first note that our variational upper bounds are
always above the corresponding Faddeev eigenvalues.  The qualitative
agreement of the first two lines hides a slight variation in the
two-body potentials chosen.  The Reid $v_8$ alone has a Faddeev
eigenvalue of $E_T = -7.59$ MeV compared to the Faddeev eigenvalue of
the RSC which is $E_T = -7.35$.  Our variational calculation of the 
Reid $v_8$ alone yields $E_T = -7.28(3)$ MeV; (accidently)  very close
to the starting point of the Faddeev calculation.  Thus, the starting
Hamiltonians of slightly different  NN potentials and the non-singular
$b$ and $d$ terms of TM(81) give the rather similar results of the top
row. Going down the columns, we see that the effect of the $a$-term
alone is to decrease the total energy by 0.11 MeV (0.13 MeV) for the
RSC/TM(81) calculation and the Reid$v_8$/TM(81) calculation
respectively.  The $c$-term has the effect of decreasing the triton
total energy by 0.72 (0.95) MeV in the two calculations.  Both the
Faddeev calculation and our present variational calculation are in
qualitative agreement and nearly quantitative agreement for the model
three-body forces which do not include the  short-range -- $\pi$-range
TBF from the $c$-term.   On the other hand, the third and fourth rows
of Table 1 {\em do}  include the  short-range -- $\pi$-range TBF from
the $c$-term. They have a discrepancy of about 0.3 MeV (three times
larger than that of the non-singular Hamiltonians)  and the
discrepancy would be even more if the NN potentials were the same. This
comparison suggests that our variational wave function is adequate for
qualitative  conclusions  about TBF's of $\pi$-range -- $\pi$-range,
but that it  cannot accurately evaluate the  short-range -- $\pi$-range
TBF and a more sophisticated~\cite{APW_95,Wir_91} variational trial
function, beyond the scope of this work,  is needed for this nearly
singular term.

Now we introduce the strength constants $a'$, $b$, and $d$  which follow from 
 the employed on-mass-shell invariant amplitudes of $\pi$N
scattering.  The  invariant amplitudes $\bar F^{+}(\nu, t)$ and 
$\bar B^{-}(\nu, t)$ are given in units
of the charged pion mass (139.6 MeV).  The potentials, however, use an isospin
formalism instead of charge states so it would seem natural to employ the
SU(2) average pion mass  $(2 m_{\pi^{+}} + m_{\pi^{0}})/3 = 138$ MeV in the
propagators and form factors and therefore to convert the quoted values
used in Eqs.  (\ref{acoef})--(\ref{ccoef}) to these units.  The results
are given in Table 2.  There is a rather dramatic change in the isospin
even coefficients between the
top two rows labeled (93) and the bottom two rows labeled (99),
reflecting the difference between the invariant amplitudes from pre-meson
factory data~\cite{hohlerbook} and the meson factory data~\cite{KH}. 
The recent invariant amplitudes  allow further tests of the PCAC/current
algebra models which underlie the Tucson-Melbourne TBF. This aspect of
the new data analyses is discussed in Ref.~\cite{Praguelec};  here we confine
ourselves to reworking the Tucson-Melbourne TBF with the new on-shell
numbers.

The rows of the Table labeled TM(93) and TM$'$(93) are obtained by
inserting  
\begin{equation}
\bar F^{+}(0,m^2_{\pi^+})
= - 0.28 m^{-1}_{\pi^+} \hspace{1cm} 
\bar F^{+}(0,2m^2_{\pi^+})\approx \sigma/f_{\pi}^2=1.03 m^{-1}_{\pi^+}
\end{equation} 
into Eqs.  (\ref{acoef})--(\ref{ccoef}), and using the old value of 
$g^2 = 179.7$; all values obtained  from the  phase shift analysis
known as KH80~\cite{hohlerbook}. One obtains $d  = -{\textstyle
\frac{g^2}{4m^3}} - \bar B^{-}(0,0)/2m$ taking $\bar B^{-}(0,0)$ either
from the model~\cite{CP_93} amplitude or the the empirical amplitude.
They are the recommended values available  in 1993~\cite{CP_93}.  Two
new invariant amplitude analyses of the meson factory data base
paramaterized as the SP98 $\pi$N phase shifts are in good
agreement~\cite{Praguelec} and we choose Ref.~\cite{KH} for the TBF
force models labeled TM(99) and TM$'$(99).  The relevant on-shell
numbers from that recent analysis are
\begin{equation}
\bar F^{+}(0,m^2_{\pi^+})
= - 0.05 \pm 0.05  m^{-1}_{\pi^+} \hspace{1cm} 
\bar F^{+}(0,2m^2_{\pi^+})\approx \sigma/f_{\pi}^2=1.40\pm 0.25  m^{-1}_{\pi^+}
\end{equation} 
with $\bar B^{-}(0,0) \approx 8.6 m^{-2}_{\pi^+}$, not much changed from
Appendix A of Ref.~\cite{Coon_79}.  The current value of $g^2 = 172.1$ was
input into the interior dispersion relation analysis of Ref.~\cite{KH}
and is therefore used in TM(99) and TM$'$(99).

\begin{table}[hbt]
\begin{center}
\begin{tabular}{ccccc}
       &  $\mu a'$   &  $\mu^{3} b$ & $\mu^{3} c$ & $\mu^{3} d$  \\ \hline
TM(81)  & -0.84      &  -2.49       &  0.98      &  -0.72   \\
TM$'$(81)  & -0.84     &  -2.49      &  0     &  -0.72   \\
TM(93)  & -0.74      &  -2.53       &  0.88      &  -0.72   \\
TM$'$(93)  &  -0.74    &   -2.53     &   0  &  -0.72   \\
TM(99) & -1.12    & -2.80      &  1.25      &  -0.75  \\
TM$'$(99) & -1.12   &  -2.80    &   0     & -0.75 \\
\label{tab2}
\end{tabular}
\caption{Expansion coefficients of the Tucson-Melbourne
$\pi-\pi$ force for $\Lambda = 5.8\mu$. 
Units of  SU(2) average pion mass $\mu$ = 138.0 MeV. The $c$ coefficient
multiplies a short-range -- $\pi$-range three-body force now known to be
spurious.}
\end{center}
\end{table}

We follow tradition and calculate the triton properties with the cutoff
in the form factor $\Lambda/ \mu = (4.1,5.8,7.1)$.  In the publications
of the Tucson-Melbourne group $\Lambda = 5.8 \mu$ has been recommended
to match the Goldberger-Treiman discrepancy~\cite{ffrefs}, another
measure of chiral symmetry breaking~\cite{Praguelec}.  The value 
$\Lambda = 7.1 \mu$ matches  the Goldberger-Treiman discrepancy $\Delta
= 0.02$ of the recent determinations of the $\pi$NN coupling constant
$g^2 = 172.1$~\cite{Praguelec}.  We don't know the reason  others have
chosen $\Lambda = 4.1 \mu$ as a test case but adopt it anyway.  Please
notice from Eq. (\ref{ccoef}) that $c$, and therefore $a'$, changes
with different values of $\Lambda$.  From Eq. (\ref{ccoef}), we see that
$\mu^2 c = \sigma/{f_{\pi}}^2 (1 + F'_{\pi NN}(0) ) - \mu^2{g^2}/{4m^3}
= \sigma/{f_{\pi}}^2(1 + \Delta (1-\Delta) )- \mu^2{g^2}/{4m^3}\approx
\sigma/{f_{\pi}}^2(1 + \Delta) - \mu^2{g^2}/{4m^3} $, because the value 
of   $\Delta= \mu^2/\Lambda^2$ varies only between $\Delta=0.06$ 
and $\Delta=0.02$ as 
$\Lambda/ \mu = (4.1,5.8,7.1)$ in Eq. (13).  Thus, the dependence of the
value of
$a'$ with $\Lambda/ \mu$ is slight, compared with the overall effect of 
the cutoff on the $\pi-\pi$ force.

The results of our calculations are presented in Figure 1 as the open
circles and open squares.  The plotted points include Monte Carlo error
bars and the lines through the symbols are drawn to guide the eye.  The
open circles show the calculated triton binding energy with Reid
$v_8$/TM$'$(93) which has  no short-range -- $\pi$-range term and the
strength constants taken from twenty year old $\pi$N scattering data.
The open squares indicate the results with the same NN potential and
the updated strength constants of TM$'$(99). Each calculation was made
variationally with the full Hamiltonian with strength constants shown
in Table 2. We indicate our calculated value of the binding energy of
the triton with the Reid $v_8$ alone ($E_T = -7.28(3)$ MeV) by a
horizontal (sparse) dotted line and the  Faddeev eigenvalue  ($E_T =
-7.59$ MeV) by the horizontal (dense) dotted line.   Our variational
upper bounds are always above the corresponding Faddeev eigenvalues.

We compare our results with calculations in the literature with the old
TBF TM(81), where the lack of a prime means that the short-range --
$\pi$-range term is {\em included}.  We do not present our own 
variational estimates with this short-range --$\pi$-range term included
as they  do not reflect the true situation (see discussion of  Table 1).
 The results of the combination
RSC/TM(81) for the three cutoffs~\cite{Chen}  (already quoted in Table
1 for the cutoff $\Lambda/\mu = 5.8$) are given by the points with an
$*$. 
%The scale of the discrepancy at ($\Lambda = 5.8 \mu$) between our
%variational calculation with  Reid $v_8$ plus TM$'$(93) given by the
%open circle  and the Faddeev calculation with RSC plus
%TM(81)~\cite{LANL_88} indicated by the $*$  can be compared with the
%scale given by the vertical distance between the two horizontal lines
%of variational versus Faddeev Reid $v_8$ alone.  
Another Faddeev
evaluation~\cite{Han_95} of the same Hamiltonian (RSC/TM(81)) is shown
as stars at the three values of $\Lambda/\mu$ and the short dashed line
interpolates between the calculated values.

  \setcounter{figure}{0}
\begin{figure}[htpb]
\vspace{1.0in}
\unitlength1.cm
\begin{picture} (5,9)(-15,2.3) 
\includegraphics{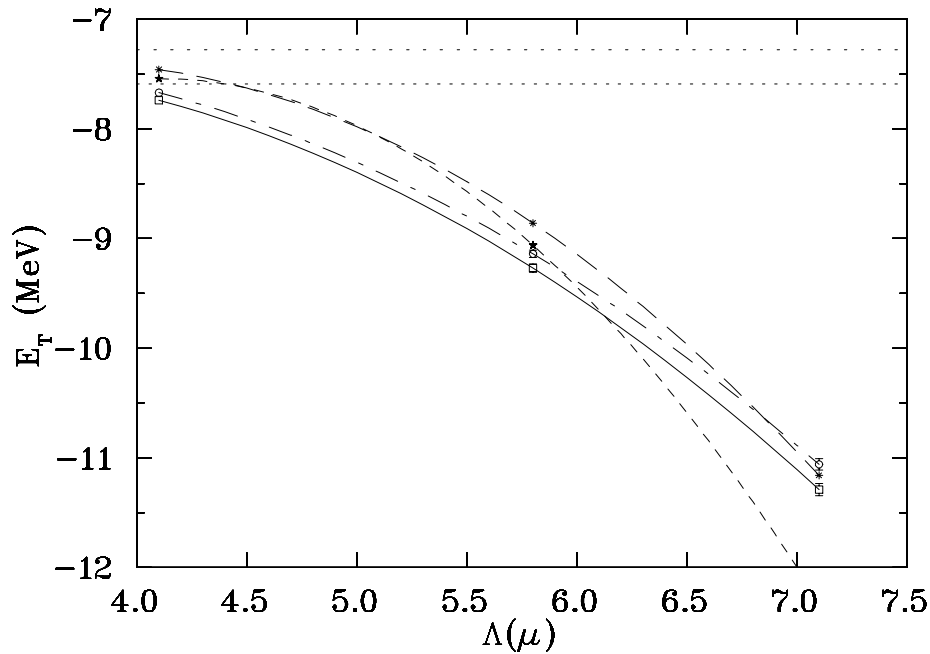}
\end{picture}
\caption{Dependence of calculated triton binding energies on $\Lambda$
for the three-body force models  
TM$'$(93)(open circles) and TM$'$(99)(open
squares).  The NN potential was the Reid $v_8$ potential.
Horizontal lines are the calculated value without a three-body force.
Two Faddeev calculations with the NN/TBF combination RSC/TM(81) are
shown for comparison.  See text for details}
\end{figure}

%\pagebreak

The models TM$'$(93) and TM$'$(99) with the spurious short-range --
$\pi$-range TBF removed (open circles and open squares) give very
similar binding energies in our calculation.  The updating of the
strength constants seems to have very little effect on the three
nucleon bound state, once the spurious term is removed.  It is
difficult to estimate the effect of removing the  short-range --
$\pi$-range force on the binding energy with the results available in
Figure 1, because both the NN potential (Reid $v_8$ versus RSC) and the
TBF (TM(81) and TM(93)) are slightly different. However, once this
spurious force is removed the two models TM$'$(93) and TM$'$(99) 
 have a similar dependence upon $\Lambda$; those two curves are
shifted vertically only slightly.  It is noteworthy that the dependence
upon $\Lambda$ is greater if the spurious  short-range -- $\pi$-range
term is included in the TBF~\cite{Chen}; and significantly greater for
the momentum space calculations of Ref. \cite{Han_95}.  One would
expect this as $\Lambda$ increases and the singular term (in one NN
separation) becomes more like a delta function.  It is a nice feature
that removal of the spurious term makes the Tucson-Melbourne two-pion
exchange force less sensitive to the cutoff.

%\pagebreak

\begin{acknowledge}
This work  was supported  by NSF grant PHY-9722122.
\end{acknowledge}

\end{document}